\documentclass[twocolumn,superscriptaddress,footinbib,showpacs,floatfix]{revtex4-1}

\usepackage{graphicx}
\usepackage{dcolumn}
\usepackage{bm}
\usepackage[usenames,dvipsnames]{xcolor}
\usepackage[colorlinks=true,urlcolor=blue,citecolor=blue, linkcolor = blue]{hyperref}
\usepackage{amssymb,amsmath}
\usepackage[percent]{overpic}
\def\be{\begin{equation}}
\def\ee{\end{equation}}
\def\ba{\begin{array}{lll}}
\def\ea{\end{array}}
\def\ber{\begin{eqnarray}}
\def\eer{\end{eqnarray}}
\begin{document}
\title{Phase-dependent noise in Josephson junctions}
\author{Forrest Sheldon}
\email{fsheldon@physics.ucsd.edu}
\affiliation{Department of Physics, University of California, San Diego, La Jolla, CA 92093, USA}
\author{Sebastiano Peotta}
\email{sebastiano.peotta@aalto.fi}
\affiliation{COMP Centre of Excellence, Department of Applied Physics, Aalto University School of Science, FI-00076 Aalto, Finland}
\author{Massimiliano Di Ventra}
\email{diventra@physics.ucsd.edu}
\affiliation{Department of Physics, University of California, San Diego, La Jolla, CA 92093, USA}

\begin{abstract}
In addition to the usual superconducting current, Josephson junctions (JJs) support a phase-dependent 
conductance related to the retardation effect of tunneling quasi-particles. This introduces a dissipative current with a memory-resistive (memristive) character and thus should also affect the current noise. By means of the microscopic theory of tunnel junctions we compute the complete current autocorrelation function of a Josephson tunnel junction and show that this 
memristive component gives rise to a non-stationary, phase-dependent noise. As a consequence, dynamic and thermal noise necessarily  
show a phase dependence otherwise absent in nondissipative JJ models. This phase dependence may be realized experimentally as a hysteresis effect if the unavoidable time averaging of the experimental probe is shorter than the period of the Josephson phase.
\end{abstract}
%
%
\maketitle

\section{Introduction}\label{sec:introduction}

The Josephson junction (JJ)~\cite{Josephson:1962,Barone_book} is the basic circuit element of superconducting electronics. Formed by a tunneling barrier between two superconductors, its primary feature is the nondissipative supercurrent $I_S = I_c\sin \gamma(t)$ (Josephson current), where $I_c$ is the critical current and $\gamma(t)$ is the gauge-invariant phase difference between the order parameters of the two superconducting electrodes~\cite{Tinkham_book}.  In Josephson's original work~\cite{Josephson:1962} it was shown that in addition to the supercurrent a JJ supports a phase-dependent and dissipative current $I_M = G(\gamma)V$, with $V$ the voltage drop across the junction.

\begin{figure}
\begin{center}
\includegraphics[width=8.6cm]{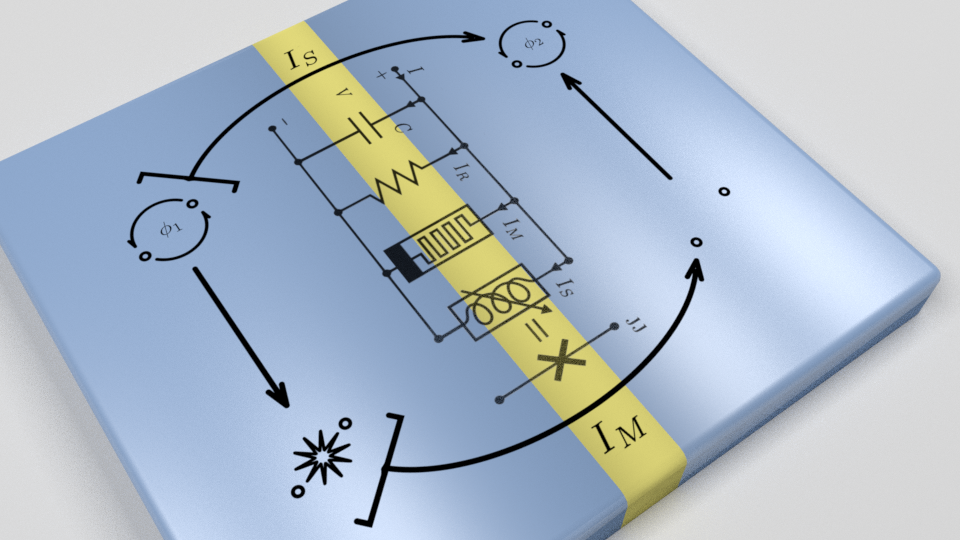}
\caption{A schematic representation of a JJ illustrating the two pair tunneling processes and the RSJ model equivalent circuit. Pairs in the left electrode with phase $\phi_1$ may tunnel to the second either directly as shown by the upper arrow ($I_S$) or by breaking, tunneling as separate quasiparticles and re-pairing with phase $\phi_2$, as shown by the lower arrow ($I_M$).  In the low voltage/frequency limit, the TJM model is well approximated by the RSJ model given in equation~(\ref{eqn:RSJ}) and shown across the tunneling barrier (yellow).  This consists of a displacement current, $C\frac{dV}{dt}$, resistive current $I_R$, supercurrent $I_S$ and a memristive component $I_M = \varepsilon G_LV \cos\gamma$.\label{fig:equiv_circuit}}
\end{center}
\end{figure}

This phase-dependent conductance (PDC) $G(\gamma) \sim\cos\gamma$  is often referred to as the `$\cos$' term and arises from the imaginary part of the superconducting response function~\cite{Josephson:1962, Barone_book, Likharev_book}.  As such, it has been interpreted as a consequence of the retarded phase-current response~\cite{Harris:1974,Harris:1975,Harris:1976,Zorin:1979}, accounting for the finite response time of the junction, or as an interference effect between quasiparticle and Cooper pair currents~\cite{Tinkham_book,Josephson:1962}.  A microscopic interpretation is supplied by the second-quantized form of the response~\cite{Stephen:1969, Barone_book} which shows the breaking, tunneling, and subsequent re-formation of a pair.  These two pair tunneling processes are illustrated schematically in Fig.~\ref{fig:equiv_circuit}.  Irrespective of interpretation, it is a memory resistive (memristive) component since it gives rise to hysteretic behavior under specific driving conditions~\cite{Peotta:2014}. 

BCS theory cannot account for the measured value of the PDC in tunnel junctions~\cite{pfl,smp}, point contacts~\cite{vd,rd} or weak links~\cite{nw,fpt,Likharev:1979}. Several effects may account for such discrepancy (see, e.g., Ref.~\cite{Zorin:1979}), but so far very few studies 
have been carried out to unravel the consequences of such memristive component. 
Recently, the PDC has been discussed theoretically in Refs.~\cite{Catelani:2011a,Leppakangas:2011}
and studied in an experiment on fluxonium qubits~\cite{Pop:2014} aimed at understanding quasiparticle-induced decoherence in superconducting qubits~\cite{Moji:1999,Lutchyn:2007,Makhlin:2001,Martinis:2009,Lenander:2011,Catelani:2011b,Nori:2005,Wilhelm:2008,Devoret:2013}.  In a recent publication~\cite{Peotta:2014}, two of us (SP and MD) have proposed a two-junction interferometer to isolate the PDC from the nondissipative pair current, allowing for a more detailed study of its properties and extraction of its hysteretic features. The question then naturally arises as to whether other fundamental properties of JJs are affected by this memristive component. 

The fluctuation-dissipation theorem suggests that the PDC should lead to a similar phase-dependent current noise.  With the possibility of isolating the PDC and the accessibility of electronic measurements reaching into the Josephson frequency range, such non-stationary noise processes may become important for the interpretation of experiments and technological applications.  For example, in the interferometer mentioned above measurement of phase-dependent noise would allow the determination of the junction phase without access to the supercurrent or applying a voltage, thus providing an avenue for nondestructive readout of the device state. The question is what type of phase-dependent noise such memristive component would induce in JJs, and how best it can be experimentally 
observed. 

In this article, by means of the microscopic theory of tunnel junctions we compute the complete current autocorrelation function of a Josephson tunnel junction.  The resulting function contains a modulation which, in appropriate limits, takes a form  $\propto\cos \gamma(t)$.   We pay particular attention to the effects of corrections to the BCS result on the subgap current response.  The correction to the response functions, introduced to match the experimentally observed broadening of the Riedel peak, also affects the subgap current response and magnitude of the noise.  We demonstrate that the expected noise variation due to the phase-dependent dissipative current is comparable to the averaged noise present at frequencies below $V_g = \frac{2\Delta}{e}$ and thus we expect it to be detectable in experiments.

That such a modulation exists has been noted previously in Ref.~\cite{Rogovin:1974} but in subsequent considerations the time average of the spectrum has been taken and the phase dependence was assumed to vanish.  However, in the thermal limit of the noise the phase of the junction may be kept stationary and thus 
it does not vanish even in a time average. Phase-dependent thermal noise has also been predicted in quantum point contacts~\cite{Martin-Rodero:1996} although the form is quite different from that expected in tunnel junctions. In Ref.~\cite{Martin-Rodero:1996} the dominant contribution to the noise comes from bound states whose energies lie within the gap and we suspect a similar role may be played by the sub-gap currents due to impurities in a tunnel junction.

In biased junctions, due to the unavoidable time averaging, a phase-dependent power spectrum cannot be defined for frequencies less than $\omega_J$. We demonstrate instead that a phase dependence can still be expected in the limit $\omega > \omega_J$.  Here we distinguish between the subgap, $\omega < \omega_g$, and quantum noise, $\omega > \omega_g$, regimes and calculate the expected phase dependence in each.  As the Josephson frequency is in the GHz range, measurement of this phase dependence will require both high-frequency and short-time resolution.
 Experimental systems designed for probing the quantum noise limit of mesoscopic systems have reached frequency-resolved measurements on the order of 100 GHz~\cite{Deblock:2003} and could potentially be adapted to the detection of such nonstationary processes.

\section{Phase-Dependent Conductance}\label{sec:pdc}

 \begin{figure}
\begin{center}
\includegraphics[width=8.6cm]{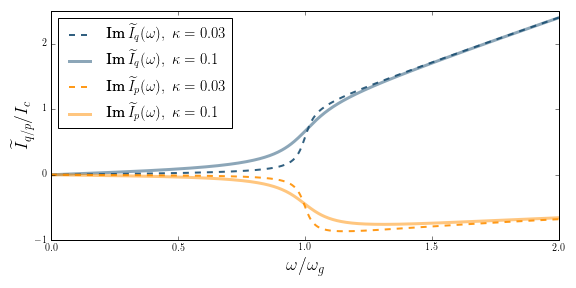}
\caption{The Fourier transforms $\text{Im}\,\widetilde{I}_p(\omega)$ and $\text{Im}\,\widetilde{I}_q(\omega)$ of the pair and quasiparticle response functions at $T=0$ given by  equations (\ref{eqn:Ip}) and (\ref{eqn:Iq}).  They are plotted in units of the gap frequency $\omega_g = \frac{2\Delta}{\hbar}$ and the critical current $I_c$ for $\kappa=\frac{\tau_g}{2\tau_r}=0.03, 0.1$.  Note that the current response below the gap frequency is enhanced with $\kappa$.\label{fig:response_functions}}
\end{center}
\end{figure}

In order to demonstrate the existence of phase-dependence noise in JJs we make use of the tunnel junction microscopic (TJM) model~\cite{Likharev_book} with a phenomenological factor which smoothes the energy gap edge.  This correction accounts for several deviations from BCS theory observed in experiments, and also produces a linear dependence of the subgap current on the smoothing parameter $\kappa = \frac{\tau_g}{2\tau_r}$.  We thus expect that the magnitude of the PDC and resulting noise will be strongly influenced by the detailed structure of the junction.

The dynamics of a generic low-transparency JJ are well described by second-order perturbation theory in the tunneling matrix elements resulting in the tunnel junction
microscopic (TJM) model where the total current $I =  I_{pair} + I_{qp}$ is the sum of the pair current $I_{pair}$ and quasiparticle current $I_{qp}$ given by,
\begin{align}
I_{pair}(t) &= \int _{-\infty}^t dt' I_p(t-t')\sin\left(\frac{\gamma(t) + \gamma(t')}{2}\right)\label{eqn:Ipair}\\
I_{qp}(t) &= \int _{-\infty}^t dt' I_{q}(t-t')\sin\left(\frac{\gamma(t) - \gamma(t')}{2}\right).\label{eqn:Iqp}
\end{align}

The time-dependent phase $\gamma(t)$ results from a voltage drop across the junction,
\begin{equation}
\frac{d\gamma}{dt} = \frac{2e}{\hbar} V(t).
\end{equation}
The material properties of the junction and superconducting electrodes are represented in the pair and quasiparticle response functions $I_p(t), I_q(t)$, respectively. We choose the form given by BCS theory at $T=0$ and with superconducting energy gaps $\Delta_1 = \Delta_2 = \Delta$.  In this case, the response functions have closed form,
 \begin{align}
  I_p(t) & = -\frac{2 I_c}{\tau_g} J_0 \left(\frac{t}{\tau_g}\right) Y_0\left( \frac{t}{\tau_g}\right) \exp\left(-\frac{t}{\tau_r}\right) \label{eqn:Ip}\\
  I_q(t) & = \frac{2 I_c}{\tau_g} J_1\left(\frac{t}{\tau_g}\right) Y_1\left( \frac{t}{\tau_g}\right) \exp\left(-\frac{t}{\tau_r}\right) - \frac{\hbar G_N}{e}\delta'(t). \label{eqn:Iq}
 \end{align}
 where $J_n, Y_n$ are the Bessel functions of the first and second kind, $\delta'$ is the derivative of the delta function, $\tau_g = \frac{\hbar}{\Delta}$ the gap timescale,  and we have included the phenomenological factor $\exp(-t/\tau_r)$ mentioned above which cuts off the algebraic decay of the Bessel functions at $t > \tau_r$~\cite{Likharev_book}.

For our purposes it will be sufficient to consider a constant voltage so that the presence of the PDC $\propto \cos \gamma(t)$ may be explicitly shown.  The phase advances linearly in time, $\gamma(t) = \omega_J t + \gamma_0$ where we have defined the Josephson frequency $\omega_J = \frac{2eV}{\hbar}$.  The junction current may then be written as
\begin{align}
I(t) = &\text{Re} \,\widetilde{I}_p \left(\frac{\omega_J}{2}\right) \sin \gamma(t) - {} \nonumber \\
& \text{Im}\,\widetilde{I}_p \left(\frac{\omega_J}{2}\right) \cos \gamma(t) +
\text{Im}\,\widetilde{I}_q \left(\frac{\omega_J}{2}\right) \label{eqn:cos_term}
\end{align}
where $\widetilde{I}_{p/q}$ are the Fourier transforms of the response functions.  These may be evaluated in terms of elliptic integrals and have been plotted in Fig.~\ref{fig:response_functions} for several values of $\kappa=\frac{\tau_g}{2\tau_r}$.  The range $\kappa = 0.03-0.1$ gives reasonable values of the peak broadening and we note that the response for $\omega < \omega_g$  of $\text{Im}\, \widetilde{I}_{p/q}$ increases with $\kappa$.  Noting that the imaginary parts of the response functions are odd in $\omega_J$ it is convenient to define the conductances $\text{Im}\,\widetilde{I}_q(\frac{\omega_J}{2}) = \sigma_0(V, T)V$ and $\text{Im}\,\widetilde{I}_p(\frac{\omega_J}{2}) = \sigma_1(V, T)V.$  From Figure~\ref{fig:response_functions} we see that for voltages/frequencies small compared to the gap voltage/frequency these are well approximated by constants, and we can define the leakage conductance $G_L = \sigma_0(V\to 0, T)$ and the ratio $\varepsilon = \lim_{V\to 0} -\frac{\sigma_1(V, T)}{\sigma_0(V, T)}$. Including the effects of a finite capacitance and fluctuations, the total junction current may then be written as,
\begin{equation}\label{eqn:RSJ}
I(t) = C\frac{dV}{dt} + G_L V (1 + \varepsilon \cos\gamma(t)) + I_c \sin \gamma(t) + I_F(t)
\end{equation}
where we have suppressed the dependence of the conductances on the temperature $T$, and regularization $\kappa$. The equivalent circuit to equation~(\ref{eqn:RSJ}) is given in Fig.~\ref{fig:equiv_circuit} and the TJM model is thus well approximated by the resistively shunted junction model (RSJ)~\cite{Likharev_book} with the phase dependent conductance $G_L(1 + \varepsilon \cos \gamma).$

The ratio $\varepsilon$ has been investigated in a number of experiments on tunnel junctions \cite{pfl, smp}, weak links \cite{vd, rd}, and point contacts \cite{nw, fpt}, consistently finding $\varepsilon \sim -1$ at low temperatures in disagreement with BCS theory which predicts $\varepsilon > 0$.  This discrepancy may be accounted for by including frequency broadening in the BCS result \cite{Likharev_book} as we have done with the exponential factors in equations (\ref{eqn:Ip}) and (\ref{eqn:Iq}).  This may be physically attributed to a finite quasiparticle lifetime, gap anisotropy and renormalization~\cite{Barone_book}. The resulting sign and magnitude of $\varepsilon$ varies from -1 to 1 depending on the specific form of the regularization.  The regularization scheme used here gives $\varepsilon \approx -\frac{1}{3}$ and displays a weak dependence on $\kappa$.  We interpret this to mean that the specific microscopic details of the junction may exert a strong influence on the quasiparticle current, as can be seen in the enhanced subgap response in Fig.~\ref{fig:response_functions}.  While the particular sign and value of $\varepsilon$ are not essential for our results, we emphasize that both theory and experiment place the magnitude of the phase dependence to be on par with the dissipative current itself.

\section{Fluctuations}\label{sec:fluct}

The PDC should provide a contribution to the current fluctuations according to the fluctuation-dissipation theorem. In fact the autocorrelation function of the noise current $I_F(t)$ can be calculated from the microscopic theory in the case of an arbitrary phase dynamics $\gamma(t)$~\cite{Zorin:1981,Likharev_book}. In the simple case of DC voltage bias the autocorrelation function of the noise current reads
\begin{equation}\label{eqn:full_noise_spectrum}
\begin{split}
\langle I_F(t) &I_F(t')\rangle_S = \frac{e}{4\pi}\int_{-\infty}^{+\infty}d\omega\, e^{i\omega(t-t')}\coth\frac{\hbar(\omega+\tfrac{\omega_J}{2})}{2k_\mathrm{B}T} \\
&\times\bigg[\text{Im}\,\widetilde{I}_q(\omega+\tfrac{\omega_J}{2}) + e^{-i(\gamma_0+\omega_J t')}\text{Im}\,\widetilde{I}_p(\omega+\tfrac{\omega_J}{2})\bigg] \\
&+ \left\{\begin{array}{c} \gamma_0 \to -\gamma_0 \\
\omega_J \to -\omega_J
\end{array}\right\}\,.
\end{split}
\end{equation}
where we denote the symmetrized autocorrelation function,
\begin{equation}
\langle I_F(t) I_F(t') \rangle_S = \frac{1}{2} \left \langle I_F(t) I_F(t')+ I_F(t') I_F(t)\right\rangle.
\end{equation}
As expected, the PDC does provide a contribution to the fluctuations given by the term proportional to the Fourier transform $\text{Im}\, \widetilde{I}_p(\omega)$ (see Fig.~\ref{fig:response_functions}) which is modulated by the phase factor $e^{-i(\gamma_0+\omega_J t')}$. Due to the modulating factor, the autocorrelation function is not simply a function of the time difference $t-t'$, which means that {\it the fluctuating current is not a stationary stochastic process}~\cite{vanKampen_book}. The autocorrelation function is only invariant under discrete time translations $\left\langle I_F(t) I_F(t')\right\rangle_S = \left\langle I_F(t+2\pi/\omega_J) I_F(t'+2\pi/\omega_J)\right\rangle_S$.

We characterize the noise with a quadratic time frequency representation (TFR) defined by
\begin{equation} \label{eqn:tfr_def}
TFR(\omega, t) = \int_{-\infty}^\infty \frac{d\tau}{2\pi}\, e^{-i\omega \tau} \langle x(t+\tau) x^*(t)\rangle_S
\end{equation}
which respects time and frequency shift covariance~\cite{Hlawatsch:1992}.  The averaging of the correlation function is performed for a fixed time $t$ over the thermal and quantum ensemble of states of the electrodes, avoiding the time averaging which is typically performed~\cite{Rogovin:1974}. The nonstationarity of the noise is reflected in the time dependence $t$ about which the decay of the correlation function localizes the noise statistics.  The resulting spectrum should display a phase dependence $\propto \cos \gamma(t)$ for frequencies larger than the Josephson frequency $\omega_J$.

From~(\ref{eqn:full_noise_spectrum}) the TFR reads
\begin{equation}\label{eqn:tfr}
\begin{split}
TFR(\omega, t) &= \frac{e}{4\pi}\coth\frac{\hbar(\omega+\tfrac{\omega_J}{2})}{2k_\mathrm{B}T} \\ &\times\bigg[\text{Im}\,\widetilde{I}_q(\omega+\tfrac{\omega_J}{2}) + e^{-i(\gamma_0+\omega_J t)}\text{Im}\,\widetilde{I}_p(\omega+\tfrac{\omega_J}{2})\bigg] \\&+ \left\{\begin{array}{c}
\gamma_0 \to -\gamma_0 \\
\omega_J \to -\omega_J
\end{array}\right\}\,.
\end{split}
\end{equation}
From here we examine the thermal and dynamic (``shot'') noise limits of this expression.

\subsection{Thermal Noise}

The contribution to the noise power spectrum at \textit{zero bias} given by the PDC is discussed in Ref.~\cite{Rogovin:1974}. However in this latter work the \textit{time-averaged} autocorrelation function was considered and, because of the modulating factor, the phase-dependent noise averages to zero at finite bias. 
From our expression~(\ref{eqn:tfr}), we consider the limit  $\hbar \omega,\,eV \ll k_\text{B}T$,
\begin{equation}\label{eqn:thermal}
TFR(\omega, t) = \frac{1}{2\pi} 2 k_B TG_L \left[ 1 + \varepsilon \cos \gamma(t)\right]
\end{equation}
which leads to the autocorrelation function
\begin{equation}\label{eq:autocorrelation}
\left\langle I_F(t) I_F(t')\right\rangle_S = 2 k_\text{B}T G_L(1+\varepsilon\cos\gamma(t'))\delta(t-t')\,.
\end{equation}
This limit is appropriate for $|t-t'| \gg \tfrac{\hbar}{k_\text{B}T}$.  We thus assume Eq.~(\ref{eq:autocorrelation}) to be valid for an arbitrary slowly varying function $\gamma(t)$. This is justified since in this case the system is never driven too far away from thermal equilibrium and the quasiparticle relaxation occurs on a small time scale of the order of $\tau_g = \tfrac{\hbar}{\Delta} \sim \tfrac{\hbar}{k_\text{B}T}$.

\subsection{Dynamic Noise}

\begin{figure}
\begin{center}
\includegraphics[width=8.6cm]{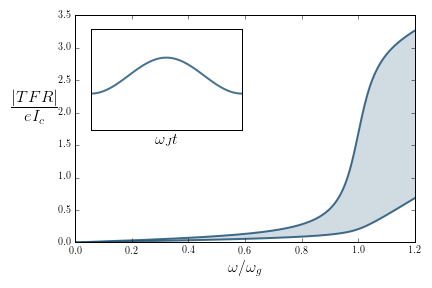}
\caption{$TFR(\omega, t)$ plotted across a period of the Josephson current.  The Josephson frequency and $\kappa$ have been chosen for the intermediate values $\omega_J = 0.005\omega_g,$ $\kappa = 0.05$.  The spectrum is plotted in units of the gap frequency $\omega_g$ and the critical current times the electron charge $eI_c$.  The inset shows the variation of the spectrum across a period at $\omega = 0.05\omega_g$, in which the $\cos\gamma(t)$ dependence is clearly visible. The functional form derived in Eq.~\eqref{eqn:TFR_dynamic} applies for $\omega \ll \omega_g$.  At frequencies above the gap, the phase dependence is strongly enhanced. \label{fig:dynamic_spectrum}}
\end{center}
\end{figure}

We now consider the noise across a biased junction.  In the thermal noise limit, the junction retains some memory of the initial conditions but over long times, the stationary phase leads to stationary noise.  In the biased case, the averaging of the noise at low frequencies limits the range in which we can observe a phase dependence.  In particular, if we consider the typical shot-noise limit $\hbar \omega_J \gg k_B T, \hbar \omega$
we find
$TFR(\omega, t) = \frac{1}{2\pi} eVG_L \left[1 + \varepsilon \cos \gamma(t)\right] $
however, the time averaging of the noise for $\omega \ll \omega_J$ makes the measurement of the phase dependence impossible at present.  
We instead consider the frequency range $\omega_J \ll \omega \ll \omega_g$ where the approximate values of $\omega_J \approx 10^9 Hz$, and $\omega_g \approx 10^{12} Hz$ place $\omega$ in the region $10^{10}-10^{11}Hz$.  Considering the limit in which $k_B T \to 0$ we may write the TFR to lowest order in $\omega_J$ as
\begin{equation} \label{eqn:TFR_dynamic}
TFR(\omega, t) = \frac{1}{2\pi} \hbar \omega G_L\left[ 1 +\varepsilon \cos\gamma(t)\, \right].
\end{equation}
While this resembles a phase dependent quantum noise, it applies in the limit $\omega \ll \omega_g$ where the current response is due to the regularization and this term should be reserved for the behavior at $\omega \gg \omega_g$ where the spectrum indeed increases linearly with $\hbar \omega$. The full spectrum has been plotted in Fig.~\ref{fig:dynamic_spectrum} across a period of the Josephson current.  Intermediate values have been chosen for the Josephson frequency $\omega_J = 0.005\omega_g,$ and  the regularization  $\kappa = 0.05$. In the inset, its variation with time shows the $\cos\gamma(t)$ dependence of Eq.~(\ref{eqn:TFR_dynamic}). The subgap noise amplitude shows a linear dependence on $\kappa$ in the range of interest.  While the form in~(\ref{eqn:TFR_dynamic}) applies for $\omega \ll \omega_g$, Fig.~\ref{fig:dynamic_spectrum} shows that in the true quantum noise limit $\omega > \omega_g$ the phase dependence of the noise is strongly enhanced.  In the conclusion, we describe an experimental arrangement to detect the phase dependence in this regime.

\section{Conclusions}\label{sec:conclusions}

We have shown that the dissipative $\cos\gamma(t)$ term in the Josephson response gives rise to non-stationary noise whose magnitude, in the thermal and dynamic regimes, displays a similar $\cos\gamma(t)$ dependence.  In an unbiased junction, this gives rise to a modulation of the normal thermal noise present at low frequencies with a magnitude of variation comparable to that of the thermal noise itself.  The presence of phase-dependent thermal noise in the junction has not yet been observed experimentally.

In a biased junction, experimental confirmation requires fluctuation measurements at very high frequency $\omega \sim 10^{10}-10^{11}Hz$ with a temporal resolution less than a period of the Josephson phase $\Delta t \sim 10^{-9}s$.  While these conditions pose an experimental challenge we do not think they are unreasonable and sketch a possible experimental approach towards their measurement. To this end, we propose an arrangement along the lines of Ref.~\cite{Deblock:2003} which utilizes an SIS junction as an on-board spectrum analyzer when capacitively coupled to the component of interest.  A JJ which intentionally contains impurities and with $\Delta_{JJ} < \Delta_{SIS}$, will exhibit fluctuations that dominate those of the SIS junction and which are modulated by the $\cos \gamma(t)$ dependence.  Through the capacitive coupling, these fluctuations in the JJ would then subject the SIS junction to microwaves which will induce quasiparticle tunneling.

The spectrum of these fluctuations may be read by biasing the SIS junction and measuring the resulting quasiparticle current. As the total current is determined by the number of microwave photons above the tunneling energy gap $ 2\Delta_{SIS} - eV_{SIS}$ the spectrum is given by the change in the quasiparticle current as $V_{SIS}$ is varied.   By tuning $V_{SIS}$ to be sensitive to photons just above the JJ gap, $\hbar\omega_g \approx 2\Delta_{SIS} - eV_{SIS}$  and modulating $V_{SIS}$ at the Josephson frequency 
$\omega_J$ over some small range, the current through the SIS detector should display hysteresis.  Finally, by taking advantage of the difference in the gaps, fluctuations above the JJ gap where quasiparticle currents and the phase dependence are enhanced, (see Fig.~\ref{fig:dynamic_spectrum}) may be measured in the subgap range of the SIS detector, thus making our predictions accessible. We then hope our findings will motivate experimental work to detect this fundamental phenomenon.

\acknowledgements{This work has been supported by DOE under Grant No. DE-FG02-05ER46204 and the Center for Magnetic Recording Research at UCSD.}

\bibliographystyle{naturemag_noURL}
\bibliography{JJ_PD_noise}

\begin{thebibliography}{10}
\expandafter\ifx\csname url\endcsname\relax
  \def\url#1{\texttt{#1}}\fi
\expandafter\ifx\csname urlprefix\endcsname\relax\def\urlprefix{URL }\fi
\providecommand{\bibinfo}[2]{#2}
\providecommand{\eprint}[2][]{\url{#2}}

\bibitem{Josephson:1962}
\bibinfo{author}{Josephson, B.~D.}
\newblock \bibinfo{title}{Possible new effects in superconductive tunnelling}.
\newblock \emph{\bibinfo{journal}{Physics letters}}
  \textbf{\bibinfo{volume}{1}}, \bibinfo{pages}{251--253}
  (\bibinfo{year}{1962}).

\bibitem{Barone_book}
\bibinfo{author}{Barone, A.} \& \bibinfo{author}{Paterno, G.}
\newblock \emph{\bibinfo{title}{Physics and applications of the Josephson
  effect}} (\bibinfo{publisher}{Wiley}, \bibinfo{year}{1982}).

\bibitem{Tinkham_book}
\bibinfo{author}{Tinkham, M.}
\newblock \emph{\bibinfo{title}{Introduction to superconductivity}}
  (\bibinfo{publisher}{Courier Corporation}, \bibinfo{year}{1996}).

\bibitem{Likharev_book}
\bibinfo{author}{Likharev, K.~K.}
\newblock \emph{\bibinfo{title}{Dynamics of Josephson junctions and circuits}}
  (\bibinfo{publisher}{Gordon and Breach}, \bibinfo{year}{1986}).

\bibitem{Harris:1974}
\bibinfo{author}{Harris, R.~E.}
\newblock \bibinfo{title}{Cosine and other terms in the josephson tunneling
  current}.
\newblock \emph{\bibinfo{journal}{Phys. Rev. B}} \textbf{\bibinfo{volume}{10}},
  \bibinfo{pages}{84--94} (\bibinfo{year}{1974}).

\bibitem{Harris:1975}
\bibinfo{author}{Harris, R.~E.}
\newblock \bibinfo{title}{Josephson tunneling current in the presence of a
  time-dependent voltage}.
\newblock \emph{\bibinfo{journal}{Phys. Rev. B}} \textbf{\bibinfo{volume}{11}},
  \bibinfo{pages}{3329--3333} (\bibinfo{year}{1975}).

\bibitem{Harris:1976}
\bibinfo{author}{Harris, R.~E.}
\newblock \bibinfo{title}{Intrinsic response time of a josephson tunnel
  junction}.
\newblock \emph{\bibinfo{journal}{Phys. Rev. B}} \textbf{\bibinfo{volume}{13}},
  \bibinfo{pages}{3818--3821} (\bibinfo{year}{1976}).

\bibitem{Zorin:1979}
\bibinfo{author}{Zorin, A.~B.}, \bibinfo{author}{Kulik, I.~O.},
  \bibinfo{author}{Likharev, K.~K.} \& \bibinfo{author}{Schrieffer, J.~R.}
\newblock \bibinfo{title}{On the sign of the quasiparticle-pair interference
  current in superconducting tunnel junctions}.
\newblock \emph{\bibinfo{journal}{Fiz. Nizk. Temp. (Sov. J. Low Temp. Phys.)}}
  \bibinfo{pages}{1138 (537)} (\bibinfo{year}{1979}).

\bibitem{Stephen:1969}
\bibinfo{author}{Stephen, M.~J.}
\newblock \bibinfo{title}{Lectures on josephson tunneling}.
\newblock In \bibinfo{editor}{Wallace, P.~R.} (ed.)
  \emph{\bibinfo{booktitle}{Superconductivity}}, vol.~\bibinfo{volume}{1},
  \bibinfo{pages}{297--326} (\bibinfo{publisher}{Gordon and Breach},
  \bibinfo{address}{New York}, \bibinfo{year}{1969}).

\bibitem{Peotta:2014}
\bibinfo{author}{Peotta, S.} \& \bibinfo{author}{Di~Ventra, M.}
\newblock \bibinfo{title}{Superconducting memristors}.
\newblock \emph{\bibinfo{journal}{Phys. Rev. Applied}}
  \textbf{\bibinfo{volume}{2}}, \bibinfo{pages}{034011} (\bibinfo{year}{2014}).

\bibitem{pfl}
\bibinfo{author}{Pedersen, N.~F.}, \bibinfo{author}{Finnegan, T.~F.} \&
  \bibinfo{author}{Langenberg, D.~N.}
\newblock \bibinfo{title}{Magnetic field dependence and $q$ of the josephson
  plasma resonance}.
\newblock \emph{\bibinfo{journal}{Phys. Rev. B}} \textbf{\bibinfo{volume}{6}},
  \bibinfo{pages}{4151--4159} (\bibinfo{year}{1972}).

\bibitem{smp}
\bibinfo{author}{Soerensen, O.~H.}, \bibinfo{author}{Mygind, J.} \&
  \bibinfo{author}{Pedersen, N.~F.}
\newblock \bibinfo{title}{Measured temperature dependence of the
  $cos\ensuremath{\phi}$ conductance in josephson tunnel junctions}.
\newblock \emph{\bibinfo{journal}{Phys. Rev. Lett.}}
  \textbf{\bibinfo{volume}{39}}, \bibinfo{pages}{1018--1021}
  (\bibinfo{year}{1977}).

\bibitem{vd}
\bibinfo{author}{Vincent, D.~A.} \& \bibinfo{author}{Deaver, B.~S.}
\newblock \bibinfo{title}{Observation of a phase-dependent conductivity in
  superconducting point contacts}.
\newblock \emph{\bibinfo{journal}{Phys. Rev. Lett.}}
  \textbf{\bibinfo{volume}{32}}, \bibinfo{pages}{212--215}
  (\bibinfo{year}{1974}).

\bibitem{rd}
\bibinfo{author}{Rifkin, R.} \& \bibinfo{author}{Deaver, B.~S.}
\newblock \bibinfo{title}{Current-phase relation and phase-dependent
  conductance of superconducting point contacts from rf impedance
  measurements}.
\newblock \emph{\bibinfo{journal}{Phys. Rev. B}} \textbf{\bibinfo{volume}{13}},
  \bibinfo{pages}{3894--3901} (\bibinfo{year}{1976}).

\bibitem{nw}
\bibinfo{author}{Nisenoff, M.} \& \bibinfo{author}{Wolf, S.}
\newblock \bibinfo{title}{Observation of a $cos\ensuremath{\varphi}$ term in
  the current-phase relation for "dayem"-type weak link contained in an
  rf-biased superconducting quantum interference device}.
\newblock \emph{\bibinfo{journal}{Phys. Rev. B}} \textbf{\bibinfo{volume}{12}},
  \bibinfo{pages}{1712--1714} (\bibinfo{year}{1975}).

\bibitem{fpt}
\bibinfo{author}{Falco, C.~M.}, \bibinfo{author}{Parker, W.~H.} \&
  \bibinfo{author}{Trullinger, S.~E.}
\newblock \bibinfo{title}{Observation of a phase-modulated quasiparticle
  current in superconducting weak links}.
\newblock \emph{\bibinfo{journal}{Phys. Rev. Lett.}}
  \textbf{\bibinfo{volume}{31}}, \bibinfo{pages}{933--936}
  (\bibinfo{year}{1973}).

\bibitem{Likharev:1979}
\bibinfo{author}{Likharev, K.~K.}
\newblock \bibinfo{title}{Superconducting weak links}.
\newblock \emph{\bibinfo{journal}{Rev. Mod. Phys.}}
  \textbf{\bibinfo{volume}{51}}, \bibinfo{pages}{101--159}
  (\bibinfo{year}{1979}).

\bibitem{Catelani:2011a}
\bibinfo{author}{Catelani, G.} \emph{et~al.}
\newblock \bibinfo{title}{Quasiparticle relaxation of superconducting qubits in
  the presence of flux}.
\newblock \emph{\bibinfo{journal}{Phys. Rev. Lett.}}
  \textbf{\bibinfo{volume}{106}}, \bibinfo{pages}{077002}
  (\bibinfo{year}{2011}).

\bibitem{Leppakangas:2011}
\bibinfo{author}{Lepp\"akangas, J.}, \bibinfo{author}{Marthaler, M.} \&
  \bibinfo{author}{Sch\"on, G.}
\newblock \bibinfo{title}{Phase-dependent quasiparticle tunneling in josephson
  junctions: Measuring the $\mathrm{cos}\ensuremath{\varphi}$ term with a
  superconducting charge qubit}.
\newblock \emph{\bibinfo{journal}{Phys. Rev. B}} \textbf{\bibinfo{volume}{84}},
  \bibinfo{pages}{060505} (\bibinfo{year}{2011}).

\bibitem{Pop:2014}
\bibinfo{author}{Pop, I.~M.} \emph{et~al.}
\newblock \bibinfo{title}{{Coherent suppression of electromagnetic dissipation
  due to superconducting quasiparticles}}.
\newblock \emph{\bibinfo{journal}{Nature}} \textbf{\bibinfo{volume}{508}},
  \bibinfo{pages}{369--372} (\bibinfo{year}{2014}).

\bibitem{Moji:1999}
\bibinfo{author}{Mooij, J.~E.} \emph{et~al.}
\newblock \bibinfo{title}{Josephson persistent-current qubit}.
\newblock \emph{\bibinfo{journal}{Science}} \textbf{\bibinfo{volume}{285}},
  \bibinfo{pages}{1036--1039} (\bibinfo{year}{1999}).

\bibitem{Lutchyn:2007}
\bibinfo{author}{Lutchyn, R.~M.} \& \bibinfo{author}{Glazman, L.~I.}
\newblock \bibinfo{title}{Kinetics of quasiparticle trapping in a cooper-pair
  box}.
\newblock \emph{\bibinfo{journal}{Phys. Rev. B}} \textbf{\bibinfo{volume}{75}},
  \bibinfo{pages}{184520} (\bibinfo{year}{2007}).

\bibitem{Makhlin:2001}
\bibinfo{author}{Makhlin, Y.}, \bibinfo{author}{Sch\"on, G.} \&
  \bibinfo{author}{Shnirman, A.}
\newblock \bibinfo{title}{Quantum-state engineering with josephson-junction
  devices}.
\newblock \emph{\bibinfo{journal}{Rev. Mod. Phys.}}
  \textbf{\bibinfo{volume}{73}}, \bibinfo{pages}{357--400}
  (\bibinfo{year}{2001}).

\bibitem{Martinis:2009}
\bibinfo{author}{Martinis, J.~M.}, \bibinfo{author}{Ansmann, M.} \&
  \bibinfo{author}{Aumentado, J.}
\newblock \bibinfo{title}{Energy decay in superconducting josephson-junction
  qubits from nonequilibrium quasiparticle excitations}.
\newblock \emph{\bibinfo{journal}{Phys. Rev. Lett.}}
  \textbf{\bibinfo{volume}{103}}, \bibinfo{pages}{097002}
  (\bibinfo{year}{2009}).

\bibitem{Lenander:2011}
\bibinfo{author}{Lenander, M.} \emph{et~al.}
\newblock \bibinfo{title}{Measurement of energy decay in superconducting qubits
  from nonequilibrium quasiparticles}.
\newblock \emph{\bibinfo{journal}{Phys. Rev. B}} \textbf{\bibinfo{volume}{84}},
  \bibinfo{pages}{024501} (\bibinfo{year}{2011}).

\bibitem{Catelani:2011b}
\bibinfo{author}{Catelani, G.}, \bibinfo{author}{Schoelkopf, R.~J.},
  \bibinfo{author}{Devoret, M.~H.} \& \bibinfo{author}{Glazman, L.~I.}
\newblock \bibinfo{title}{Relaxation and frequency shifts induced by
  quasiparticles in superconducting qubits}.
\newblock \emph{\bibinfo{journal}{Phys. Rev. B}} \textbf{\bibinfo{volume}{84}},
  \bibinfo{pages}{064517} (\bibinfo{year}{2011}).

\bibitem{Nori:2005}
\bibinfo{author}{You, J.~Q.} \& \bibinfo{author}{Nori, F.}
\newblock \bibinfo{title}{Superconducting circuits and quantum information}.
\newblock \emph{\bibinfo{journal}{Physics Today}}
  \textbf{\bibinfo{volume}{58}}, \bibinfo{pages}{42} (\bibinfo{year}{2005}).

\bibitem{Wilhelm:2008}
\bibinfo{author}{Clarke, J.} \& \bibinfo{author}{Wilhelm, F.~K.}
\newblock \bibinfo{title}{{Superconducting quantum bits}}.
\newblock \emph{\bibinfo{journal}{Nature}} \textbf{\bibinfo{volume}{453}},
  \bibinfo{pages}{1031--1042} (\bibinfo{year}{2008}).

\bibitem{Devoret:2013}
\bibinfo{author}{Devoret, M.~H.} \& \bibinfo{author}{Schoelkopf, R.~J.}
\newblock \bibinfo{title}{Superconducting circuits for quantum information: An
  outlook}.
\newblock \emph{\bibinfo{journal}{Science}} \textbf{\bibinfo{volume}{339}},
  \bibinfo{pages}{1169--1174} (\bibinfo{year}{2013}).

\bibitem{Rogovin:1974}
\bibinfo{author}{Rogovin, D.} \& \bibinfo{author}{Scalapino, D.}
\newblock \bibinfo{title}{Fluctuation phenomena in tunnel junctions}.
\newblock \emph{\bibinfo{journal}{Annals of Physics}}
  \textbf{\bibinfo{volume}{86}}, \bibinfo{pages}{1 -- 90}
  (\bibinfo{year}{1974}).

\bibitem{Martin-Rodero:1996}
\bibinfo{author}{Mart\'{\i}n-Rodero, A.}, \bibinfo{author}{Yeyati, A.~L.} \&
  \bibinfo{author}{Garc\'{\i}a-Vidal, F.~J.}
\newblock \bibinfo{title}{Thermal noise in superconducting quantum point
  contacts}.
\newblock \emph{\bibinfo{journal}{Phys. Rev. B}} \textbf{\bibinfo{volume}{53}},
  \bibinfo{pages}{R8891--R8894} (\bibinfo{year}{1996}).

\bibitem{Deblock:2003}
\bibinfo{author}{Deblock, R.}, \bibinfo{author}{Onac, E.},
  \bibinfo{author}{Gurevich, L.} \& \bibinfo{author}{Kouwenhoven, L.~P.}
\newblock \bibinfo{title}{Detection of quantum noise from an electrically
  driven two-level system}.
\newblock \emph{\bibinfo{journal}{Science}} \textbf{\bibinfo{volume}{301}},
  \bibinfo{pages}{203--206} (\bibinfo{year}{2003}).

\bibitem{Zorin:1981}
\bibinfo{author}{Zorin, A.}
\newblock \bibinfo{title}{Voltage fluctuations in josephson tunnel junctions}.
\newblock \emph{\bibinfo{journal}{Physica B+C}} \textbf{\bibinfo{volume}{108}},
  \bibinfo{pages}{1293 -- 1294} (\bibinfo{year}{1981}).

\bibitem{vanKampen_book}
\bibinfo{author}{van Kampen, N.~G.}
\newblock \emph{\bibinfo{title}{Stochastic processes in physics and chemistry}}
  (\bibinfo{publisher}{Elsevier}, \bibinfo{year}{2007}).

\bibitem{Hlawatsch:1992}
\bibinfo{author}{Hlawatsch, F.} \& \bibinfo{author}{Boudreaux-Bartels, G.~F.}
\newblock \bibinfo{title}{Linear and quadratic time-frequency signal
  representations}.
\newblock \emph{\bibinfo{journal}{IEEE Signal Processing Magazine}}
  \textbf{\bibinfo{volume}{9}}, \bibinfo{pages}{21--67} (\bibinfo{year}{1992}).

\end{thebibliography}

\end{document}